\documentclass[twocolumn]{aastex631}

\usepackage{amsmath}
\usepackage{todonotes}
\usepackage{xspace}

\newcommand\HFR{HFR\xspace}

\begin{document}

\title{Radial flow component of Sun's high frequency retrograde inertial waves}

\correspondingauthor{Chris S. Hanson}
\email{hanson@nyu.edu}

\author[0000-0003-2536-9421]{Chris S. Hanson}
\affiliation{Center for Astrophysics and Space Science, \\
NYUAD Institute, New York University Abu Dhabi,\\
Abu Dhabi, UAE}

\author[0009-0007-8208-3960]{Vivek Menon}
\affiliation{Center for Astrophysics and Space Science, \\
NYUAD Institute, New York University Abu Dhabi,\\
Abu Dhabi, UAE}

\author[0000-0003-2896-1471]{Shravan Hanasoge}
\affiliation{Department of Astronomy and Astrophysics,\\ Tata Institute of Fundamental Research,\\ Mumbai, India}
\affiliation{Center for Astrophysics and Space Science, \\
NYUAD Institute, New York University Abu Dhabi,\\
Abu Dhabi, UAE}

\author{Katepalli R. Sreenivasan}
\affiliation{Department of Physics, New York University, New York, USA}
\affiliation{Center for Astrophysics and Space Science, \\
NYUAD Institute, New York University Abu Dhabi,\\
Abu Dhabi, UAE}

\shortauthors{Hanson et al.}


\begin{abstract}
Solar inertial modes have the potential to surpass the diagnostic capabilities of acoustic waves in probing the deep interior of the Sun. The fulfillment of this potential requires an accurate identification and characterization of these modes. Among the set of detected inertial modes, the equatorially anti-symmetric ``high-frequency retrograde'' (\HFR) modes has attracted special interest because numerical studies have suggested that they are not purely toroidal, as initial observations suggested, and predicted that they would possess a significant radial flow signal at depth. Here, we analyze $\sim$13 years of HMI/SDO 5$^\circ$ ring tiles, and discover a horizontal-divergence signal, directly connected to radial flows, in the near surface layers of the Sun. We demonstrate that this signal is indeed part of the \HFR modes and not spatial leakage from prograde flows associated with magnetic regions. The amplitudes of the horizontal divergence are approximately half that associated with radial vorticity. We also report the presence of a ridge of enhanced power, although with a signal-to-noise ratio of 0.3, in the retrograde frequencies that coincides with the \HFR latitudinal overtones reported by models. Using numerical linear models we find reasonable agreement with observations, though future work on boundary considerations and the inclusion of the near-surface may improve future inferences. This is the first instance where numerical studies of solar inertial modes have guided observations, giving further confidence to past inferences that rely upon numerical models.

\end{abstract}


\section{Introduction} \label{sec:intro}

Helioseismic inferences of the solar interior have classically relied on acoustic (p) and surface-gravity (f) modes, which have pressure and gravity as their restoring forces, respectively. Recent years have seen a significant shift in focus for solar interior research towards inertial waves. Inertial waves, or oscillations, occur within rotating fluids and atmospheres where the Coriolis force acts as a restoring force. These waves typically have oscillation frequencies comparable to or lower than the rotation rate. Unlike p and f modes, which are generated and confined to the solar surface, inertial modes are thought to propagate deep in the solar interior and may be sensitive to parameters such as adiabatic gradients and viscosity. 

Inertial oscillations occur on global scales, and it is appropriate to describe them in terms of spherical harmonics $Y_\ell^m$, where $\ell\ge0$ is harmonic degree and $-\ell\le m\le\ell$ is the azimuthal order. The difference $\ell-|m|$ signifies the number of nodes in the latitude of the flow; modes with small values of $\ell-|m|$ are equatorially confined and vice versa. The terms sectoral and tesseral, often used in the literature, refer to flow patterns where $\ell =m$ or $\ell\neq m$, respectively.

\citet{loeptien_etal_2018} were the first to report the existence of hydrodynamical inertial waves in the Sun. They analyzed six years of flow maps derived from the Helioseismic and Magnetic Imager \citep[HMI/SDO]{scherrer_etal_2012,schou_etal_2012a} and discovered an unambiguous signal in the power spectrum of the surface radial vorticity. The dispersion relation (which relates length scale and frequency) of this signal follows surprisingly well the classical Rossby-Haurwitz prediction of thin-shell Rossby-wave propagation. These waves are purely toroidal, i.e. no apparent radial or divergent flow signals, sectoral ($\ell=m$ only), equatorially confined and possess lifetimes on the order of months. Due to strong agreement with theory, the modes of \citet{loeptien_etal_2018} were attributed as solar Rossby waves. 
Numerous follow-up studies have confirmed these findings using different helioseismic methods and data sets \citep{liang_etal_2019,hanasoge_mandal_2019,hanson_etal_2020,proxauf_etal_2020,hanson_hanasoge_2024}. Further analysis of the surface flow power spectra by \citet{gizon_etal_2021} revealed a number of additional isolated modes that were identified as high-latitude, critical-latitude and equatorial inertial modes. Unlike the sectoral Rossby waves, these modes do not follow clear dispersion relations. 
A notable inertial mode in this set is the isolated high-latitude $m=1$ retrograde mode that has been measured independently by numerous methods and data sets \citep{hathaway_etal_2013,gizon_etal_2021,liang_gizon_2025}, and is thought to {limit the amplitude of the} solar differential rotation \citep{bekki_etal_2024}. 

The last member of the solar inertial modes reported thus far is the high-frequency retrograde vorticity modes detected by \citet{hanson_etal_2022}, which are the subject of this paper. We refer to these anti-symmetric modes as the high-frequency retrograde (\HFR) modes because of their relatively high frequencies compared to the other inertial modes discussed previously.
The \HFR ridge comprises a distinct set of inertial modes, so far detected only in radial vorticity measurements of solar surface motions.  The vorticity signals are antisymmetric ($\ell=m+1$) across the equator and have oscillation frequencies roughly three times that of sectoral Rossby modes (at the same wavenumber).  \citet{hanson_etal_2022} reported that the motions near the solar surface were toroidal (vortical) with no detectable poloidal (divergent) flows that could be differentiated from spatial leakage of prograde signals. Spatial leakage occurs in the power spectrum of the inertial modes due to only part of the solar surface being visible to Earth. Due to the high retrograde phase speed of the \HFR modes, \citet{hanson_etal_2022} investigated the possibility that they were mixed mode-oscillations, Rossby waves altered by the influence of convection, magnetism or gravity. All three possible mixed-mode scenarios have been proposed and explored by theory \citep[see review of ][]{zaq_etal_2021}, and explain the observed high frequencies. Through observational, numerical and theoretical arguments, we had discounted these possibilities.

Recent numerical studies \citep{triana_etal_2023,bhattacharya_hanasoge_2023,bekki_2024,blume_etal_2024,jain_etal_2024} have reported the appearance of the \HFR modes in both uniformly rotating and stratified differentially rotating spheres. Most notably, some of the studies reported that the modes they identified as the \HFR modes are not purely toroidal and may possess a significant radial flow (i.e. poloidal flow) signal at depth. Recently, \citet{blume_etal_2024} reported the appearance of the \HFR modes in their nonlinear simulations and suggested that they may be mixed with prograde thermal Rossby waves. Meanwhile, \citet{bekki_2024} compared the observed frequency and line widths of \HFR modes with those predicted by an eigenmode solver, reporting that they were consistent when the bulk of the convection zone is almost neutrally buoyant or even subadiabtic. This result further constrained the estimates provided by \citet{gizon_etal_2021} who utilized high and critical latitude modes. Guided here by the predictions of \HFR radial flow components, we provide observational evidence of a surface poloidal signal, demonstrating that it is indeed a characteristic of the \HFR modes.

\section{Observations}\label{sec:observations}
Motions in the solar interior due to inertial modes result in shifts in the frequencies and eigenfunctions of the f or p normal modes, which is how we detect them. To parameterize these motions, we use the vector spherical harmonics basis, wherein a general vector flow field may be decomposed into toroidal (mass-preserving horizontal motions with zero radial velocity) and poloidal (a combination of radial and horizontal flows) components. By construction, toroidal flows have no radial flow and the divergence of the horizontal components is therefore zero. In contrast, the horizontal divergence of poloidal flows is non-zero because the radial velocity is finite. Thus, we use the horizontal divergence as a proxy for radial flows. 

\subsection{Computing surface flows}

We use the horizontal flows inferred by the HMI ring-diagram pipeline\footnote{\url{http://jsoc.stanford.edu/HMI/Rings.html}} from the 19\textsuperscript{th} May 2010 to 17\textsuperscript{th} May 2023. There are a number of detailed steps involved in the generation of horizontal flow maps through ring-diagram analysis \citep{hill_1988} and the interested reader may refer to \citet{bogart_etal_2011a,bogart_etal_2011b} for more information. In brief, the full visible disk of the Sun is captured by the HMI Doppler camera every 45 seconds, which can detect the surface oscillations caused by the internal f and p modes. A small region of this image, i.e., a small patch of the solar surface, is then tracked for a period of time (e.g., $\sim$8 hours). This region is extracted from full-disk images and combined into a 3D array $\hat{\Psi}(x,y,t)$, commonly referred to as a Doppler cube. Since the patches are small compared to the solar radius, we employ a Cartesian coordinate system, with $x$ aligned in the direction of solar rotation, $y$ in the direction of solar north and $z$ pointing outward from the surface. A 3D Fourier transform is then performed on $\hat{\Psi}$ to obtain the spectral components $\Psi(k_x,k_y,\omega)$ of the oscillations, where $k$ and $\omega$ are the wave number and temporal frequency, respectively. The complex cube is multiplied by its complex conjugate to generate the power spectrum, $|\Psi\Psi^*|$. 

In the absence of horizontal flows, the acoustic waves appear at fixed frequency $\omega$ as distinct symmetric rings in $k$ space. The presence of horizontal flows, e.g., those generated by inertial modes, breaks the symmetry of the rings by inducing measurable frequency shifts $\delta\omega$. Fits to these rings are performed for a large set of f and p modes by the HMI pipeline, obtaining the `mode fits' $U_x$ and $U_y$. Specifically,
\begin{equation}
    \delta\omega^i = k^i_xU_x + k^i_yU_y,
\end{equation}
is the frequency shift induced in mode $i$ of wavenumber $\mathbf{k}=[k_x,k_y]$ propagating through a uniform horizontal flow field.
Each f or p mode of specific wave number $|\mathbf{k}|$ propagates within a cavity near the surface, where the depth depends on the mode wavelength and frequency. The fits $U_x$ and $U_y$ are thus considered the depth-averaged horizontal flow. The flow field as a function of depth, $u_x(z)$ and $u_y(z)$, can be inferred through inverse methods \citep[e.g.,][]{pijpers_thompson_1994} using many different modes with different propagation depths. This procedure is repeated in the HMI pipeline across most of the visible disk, forming a grid of patches (or tiles) of horizontal flows across the solar disk at every time step.  

Previous inertial mode studies that have examined ring diagram flow maps have predominantly analyzed horizontal flows $u_x(z)$ and $u_y(z)$ of the 15$^\circ$ and 30$^\circ$ tiles. Here, we analyze the 5$^\circ$ tiles, which utilize high wavenumber modes that are confined closely to the surface. Unlike the 15$^\circ$ and 30$^\circ$ products, the $5^\circ$ flow maps are not available due to the large amount of associated data \citep{bogart_etal_2011b}. Here, we average the mode fits $U$ (which have the units m/s) for the f mode, p$_1$ and p$_2$ modes, for the wavelength range of $500\leq|\mathbf{k}|$R$_\odot\leq1500$. Unlike the computation of depth-localized flows $u(z)$, this approach averages out depth information. We recently demonstrated that this approach has enabled the detection of Rossby waves up to $m=30$ \citep{hanson_hanasoge_2024}, though at the cost of the ability to infer its depth structure. The f- and p-mode fits averaged here are located in a cavity {$\leq4$~Mm $=4000$~km} below the surface and thus potential inferences that may be made by performing these inversions would only be valid in the very-near surface of the Sun ($\sim0.6\%$ in radius). 

\subsection{Computing the inertial-wave spectrum}
With the averaged modes fits $\hat{U}_x$ and $\hat{U}_y$ for every patch and time step, we follow the remapping method of \citet{loeptien_etal_2018} in order to obtain horizontal-flow maps in the equatorial rotation frame $\Omega_{\rm eq}/2\pi = 453.1$~nHz.
We then compute both the horizontal divergence $\zeta$ and radial vorticity $\eta$ through
\begin{equation}
    \zeta(\theta,\phi,t)  = \frac{1}{\textrm{R}_\odot\sin\theta}\left[\partial_\theta\left(\hat{U}_\theta(\theta,\phi,t)\sin\theta\right) + \partial_\phi\hat{U}_\phi(\theta,\phi,t)\right]
\end{equation}
\begin{equation}
    \eta(\theta,\phi,t)  =  \frac{1}{\textrm{R}_\odot\sin\theta}\left[\partial_\theta\left(\hat{U}_\phi(\theta,\phi,t)\sin\theta\right)-\partial_\phi\hat{U}_\theta(\theta,\phi,t)\right], 
\end{equation}

where $\hat{U}_\phi=\hat{U}_x$ and $\hat{U}_\theta=-\hat{U}_y$ at each patch are flows in the longitude and latitudinal directions, respectively.
Spherical-harmonic and temporal-Fourier transforms are performed in order to obtain the power spectrum of the radial vorticity $P_\eta$,
\begin{equation}
    P_\eta (\ell,m,\omega) = \left|\int\limits_0^T\int\limits_0^{2\pi}\int\limits_0^{\pi}\eta(\theta,\phi,t) Y^m_\ell(\theta,\phi) e^{-im\phi+i\omega t}\sin\theta\textrm{d}\theta\textrm{d}\phi\textrm{d}t\right|^2,
\end{equation}
which may also be extended to the horizontal divergence $\zeta$ in order to obtain its power spectrum $P_\zeta$.

\subsection{Observational results}
\begin{figure*}
    \centering
    \includegraphics[width=\linewidth]{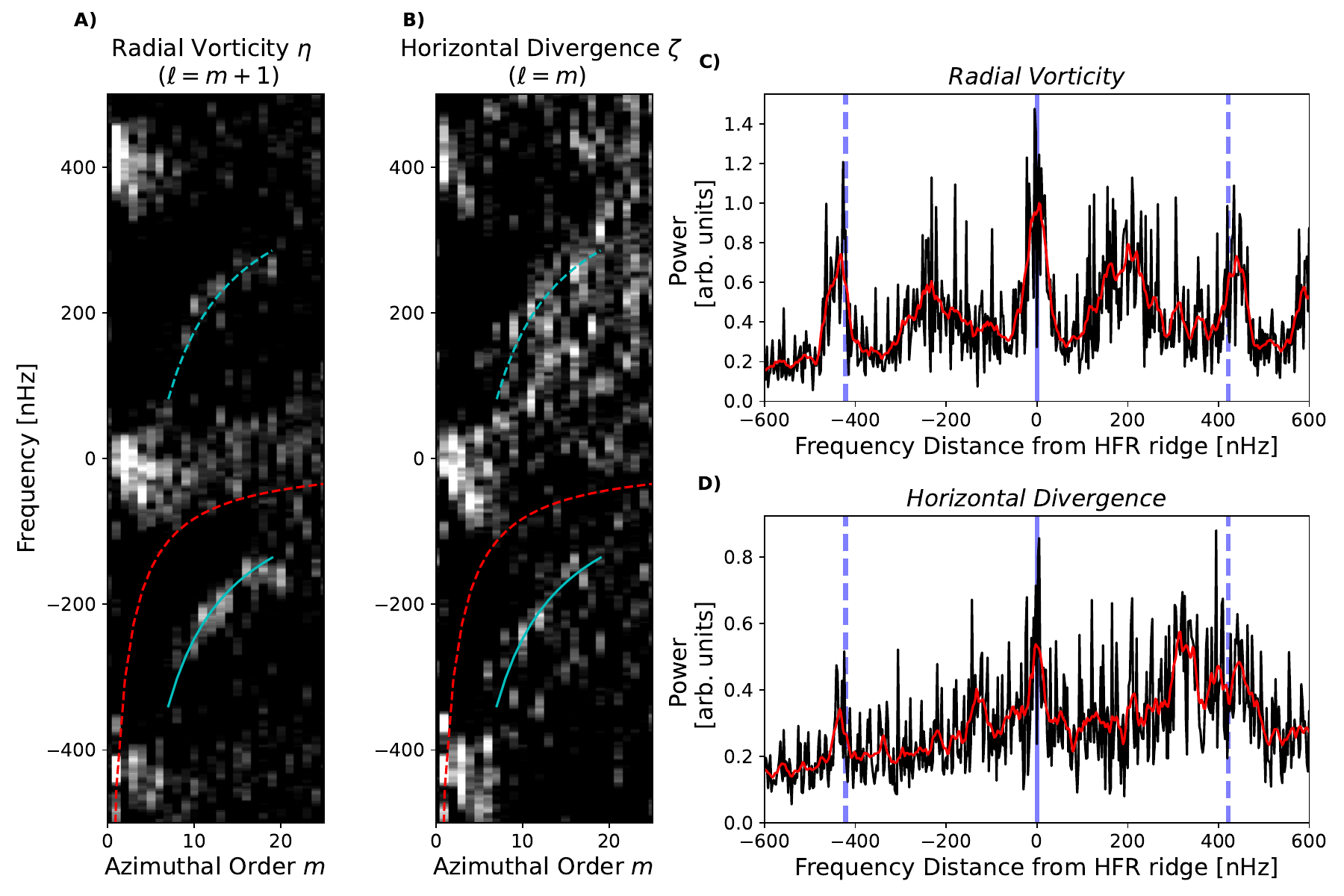}
    \caption{Smoothed $\ell=m+1$ radial-vorticity (\textbf{A}) and $\ell=m$ horizontal-divergence (\textbf{B}) power spectra for \HFR modes from the 5-deg ring-diagrams mode fits. The smoothing is for visual aid and was performed by convolving a rectangular box filter of width $\sim24$~nHz in the frequency dimension.  We have included the classical Rossby-Haurwitz dispersion (red dashed line) for context. An approximation of the \HFR dispersion and their leakage frequencies are shown by the cyan solid and dashed lines, respectively. Panels \textbf{C} and \textbf{D} show the $m$-averaged ($10\leq m \leq 15$) power-spectrum slices centered on the \HFR frequencies of radial vorticity and horizontal divergence, respectively. Both panels are normalized to the peak of the smoothed $m$-averaged vorticity spectrum. We show both the full frequency resolution (black) and smoothed spectra (red). The location of the centered \HFR power and their leakage locations are shown by the blue solid and dashed vertical lines, respectively.}
    \label{fig.spectra}
\end{figure*}

Figure~\ref{fig.spectra} panel \textbf{A} (\textbf{B}) shows power spectra of the radial vorticity (horizontal divergence) in the $\ell =m+1$ ($\ell=m$) channel. As reported by \citet{hanson_etal_2022}, there is a distinct ridge of power in the radial vorticity (panel \textbf{A}) between -100 and -300~nHz, where negative frequency implies retrograde motion, for azimuthal orders $m>7$. These are the \HFR modes. However, unlike \citet{hanson_etal_2022}, who were unable to clearly identify a divergence signal in the 15$^\circ$ maps, the $5^\circ$ flow maps reported here show signals in the horizontal divergence along the same dispersion ridge measured in the vorticity. The horizontal divergence signal is only visible in the $\ell=m$ channel, suggesting that the latitudinal eigenfunction of the divergence is symmetric about the equator and likely attains its maximum there as well.

Our previous inability to detect a divergence signal near the \HFR frequencies was a consequence of both the low signal-to-noise ratio and the spatial leakage of low frequency prograde signals seen between $0\leq\omega/2\pi\leq300$~nHz. Being able to observe only part of the solar surface from Earth, there is leakage in the power spectrum as a result of the windowing function. Specifically, a signal in $m$ with a frequency $\omega_m$ will be replicated in the $m\pm1$ channels with a diminished amplitude at $\omega_m/2\pi\mp 421.4$~nHz. {This shift in frequency is the difference between the solar equatorial rotation rate (453.1~nHz) and the Earth's (observer's) orbital period (31.7~nHz).} This was demonstrated with synthetic flow maps in \citet{liang_etal_2019} {(their Fig.~5)} and \citet{hanson_etal_2020} {(their Fig.~A1)} for time-distance and GONG ring-diagram flow maps, respectively. 
As such, any power between $0\lesssim\omega/2\pi\lesssim400$~nHz, likely due to magnetism which rotates faster than the reference frame \citep[see][]{beck_2000}, would leak down to the frequencies of $-400\lesssim\omega/2\pi\lesssim0$~nHz, where the \HFR modes are located. 
From Fig~\ref{fig.spectra}, we  see the \HFR ridge in divergent in the retrograde frequencies, but in the prograde frequencies, it is difficult to differentiate a ridge from the background. A qualitative argument for why the retrograde divergence signal belongs to the \HFR modes is that, if the signals in the retrograde frequencies were due to leakage of prograde noise, the entire prograde signal would be replicated in the retrograde, not just a few frequency bins that happen to align with the \HFR signal.

For a more quantitative assessment, we perform $m$ averaging to improve the signal-to-noise in the horizontal-divergence spectrum. We extract the $m$ slices ($10\leq m \leq 15$) of the \HFR modes reported by \citet{hanson_etal_2022}. Each slice is then translated into the frequency grid by the resonant frequency of the \HFR mode, so that the centers of all modes are aligned with zero frequency. We then average all the centered power-spectral slices. Figure~\ref{fig.spectra}\textbf{C} and \ref{fig.spectra}\textbf{D} show the $m$-averaged power spectrum of radial vorticity and horizontal divergence, respectively. Examining radial vorticity, the ridge associated with the previously reported \HFR frequencies is the dominant feature, with the replicated ridges at $\pm421.4$~nHz diminishing in amplitude. For the horizontal divergence, a Lorentzian-like profile coincides with reported \HFR frequencies, though its peak-power after accounting for background noise profiles is approximately a third of that of vorticity. A replicated peak is also present at $-421.4$~nHz. However, at $+421.4$~nHz, there are a number of peaks, some of which may be the result of prograde magnetic inflows. 

\begin{figure}
    \centering
    \includegraphics[width=\linewidth]{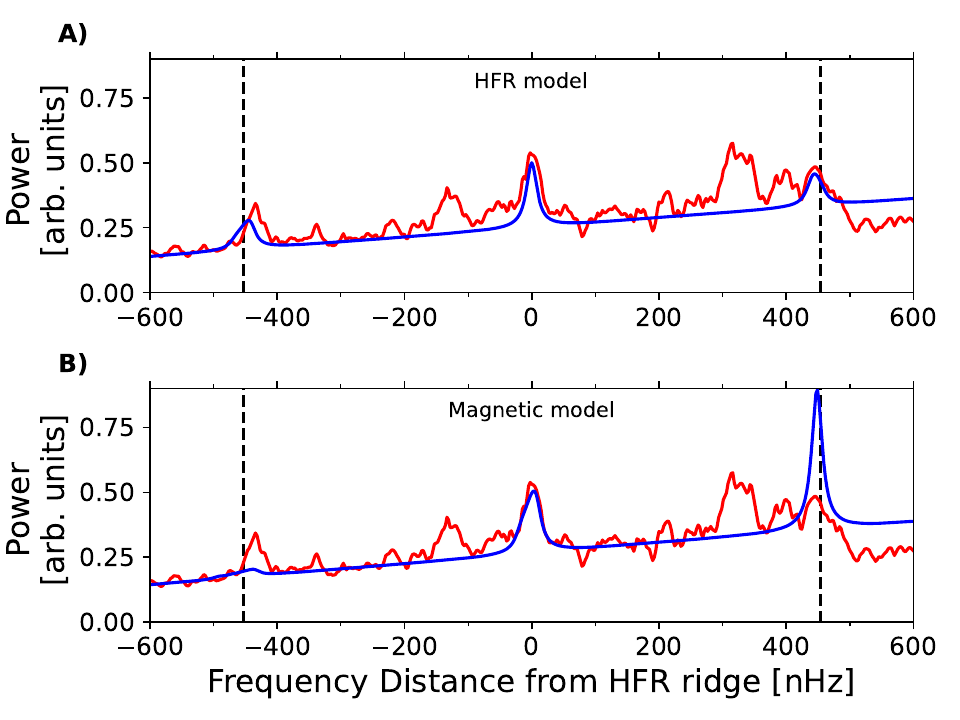}
    \caption{Comparison of synthetic $m$-averaged power spectrum  (blue lines) to the observed, smoothed $m$-averaged \HFR spectrum (red lines). Panel \textbf{A} compares the observed horizontal-divergence ($\zeta$) spectrum with a synthetic model that assumes that the $\zeta$ signal originates at the same frequencies as the \HFR observations. Panel \textbf{B} compares the observations with a synthetic model where the $\zeta$ signal originates at prograde frequencies, presumably by magnetism. The dashed lines indicate the locations of mode leakage. The observations are consistent with the model where the horizontal-divergence signal is due to the \HFR modes.}
    \label{fig.Synthetics}
\end{figure}

In order to differentiate the signal from the leakage, we compute two synthetic flow maps of inertial modes. The first model has a mode dispersion chosen to be three times that of the Rossby-Haurwitz modes, which approximates the \HFR dispersion, with a full width at half maximum (inverse lifetime) at 10~nHz. The second model has the dispersion frequencies shifted in the prograde direction by 421.4~nHz, to emulate magnetic active-region inflow signals. For each $m$, the mode amplitudes are set to unity. We then compute the inertial spectra as outlined in the previous sections and average in $m$ as well. In doing so, we quantify the amount of leakage for the 5$^\circ$ flow maps analyzed here. We fit a linear polynomial to the observed $m$-averaged spectra of horizontal divergence  (Fig~\ref{fig.spectra}\textbf{D}) in order to derive an estimate of the background power. We then scale the synthetic spectra such that the central Lorentzian is consistent with the observations. Figure~\ref{fig.Synthetics} shows the two synthetic models superimposed on the smooth $m$-averaged horizontal-divergence spectrum. In the case where frequencies of the divergence-signal are consistent with the \HFR modes, the synthetic leakage-mode amplitudes at $\pm421.4$~nHz are well matched with the observations. In the case where the signal is the result of spatial leakage from a prograde signal, the amplitudes are not consistent with the observations. These results confirm that the \HFR modes indeed possess a horizontal-divergence signal at the near surface.

We note that numerical studies have also suggested that the sectoral solar Rossby wave may be mixed with the convective prograde modes \citep{bekki_etal_2022b}, and, as such, may potentially possess a $\ell=m+1$ poloidal component. We performed the same analysis on the solar Rossby modes as we did for the \HFR modes and found no evidence, above the noise, for a surface poloidal flow component (see Fig.~\ref{fig.spectra_RH}). Given the improved signal-to-noise ratio of these data, we conclude that in the near-surface layers of the Sun the motions of the solar sectoral Rossby waves are toroidal.

\subsection{Evidence of \HFR overtones}
{The nonlinear convective simulations of \citet{blume_etal_2024} and analytical models of \citet{jain_etal_2024},} have reported the presence of a sequence of ridges, beginning with the \HFR mode, with successively higher frequencies (more retrograde). Two of these ridges were symmetric, and others were anti-symmetric with respect to the equator. \citet{blume_etal_2024} noted the first ridge, of lowest retrograde frequency and anti-symmetric across the equator, coincided with the observed \HFR frequencies of \citet{hanson_etal_2022}. \citet{blume_etal_2024} {and \citet{jain_etal_2024}} both speculated that the \HFR ridge and the other higher frequency ridges may belong to the same family, with each high frequency ridge corresponding to a greater number of nodes in the latitudinal eigenfunction.  The first symmetric ridge, which is more retrograde than the \HFR modes, was also reported by the linear study of \cite{bhattacharya_hanasoge_2023}.

\begin{figure*}
    \centering
    \includegraphics[width=\linewidth]{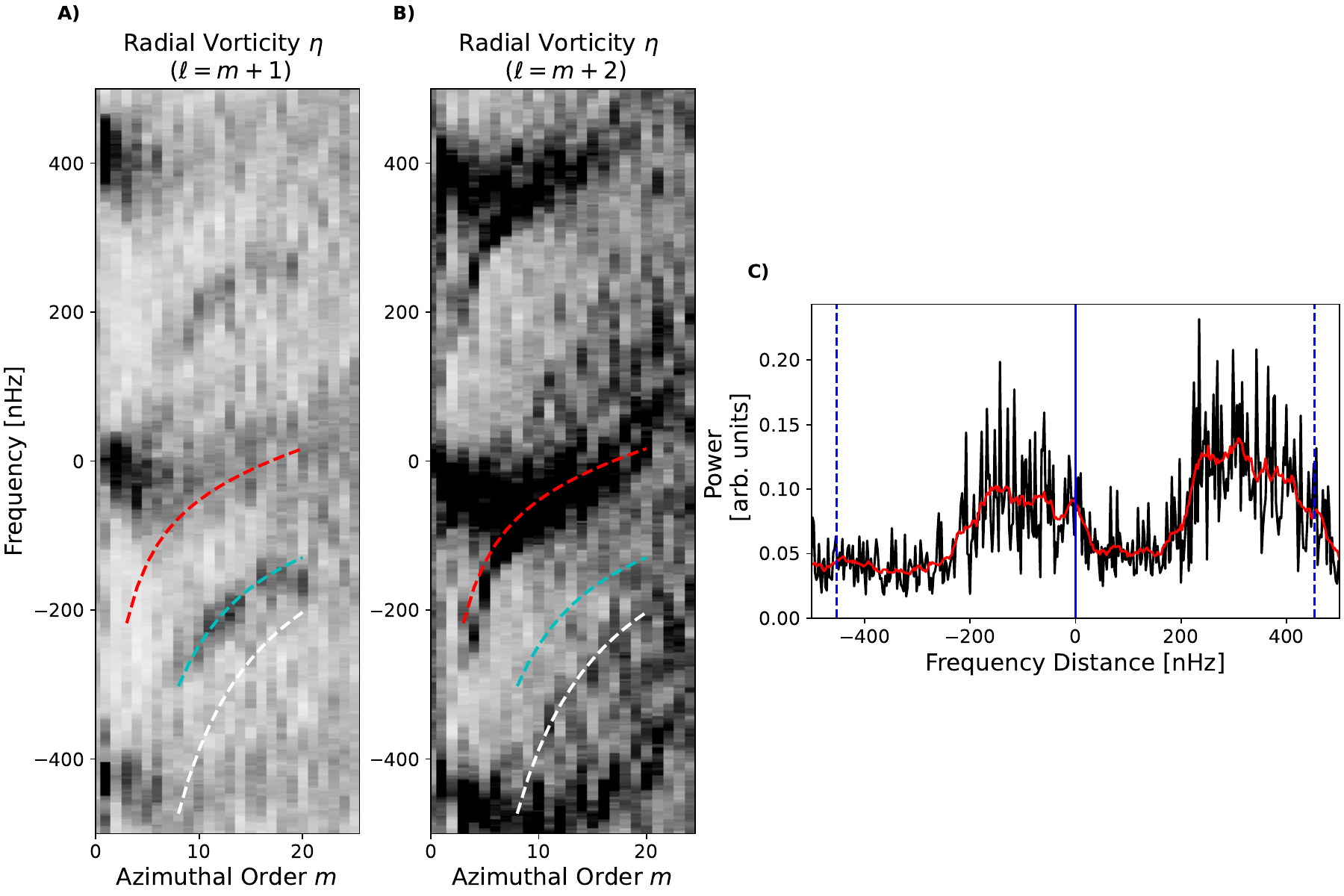}
    \caption{Radial vorticity  in the $\ell=m+1$ (\textbf{A}) and $\ell=m+2$ (\textbf{B}) channels. Same figure notation as Fig.~\ref{fig.spectra}, though we have inverted the spectrum colour table for viability. For reference we hightlight the Rossby dispersion (red dashed), the \HFR dispersion (cyan dashed) and possible location of of the \HFR overtone (white dashed). Panel \textbf{C} shows the $m$-averaged spectra of the $\ell=m+2$ (panel \textbf{B}) along the white dashed lines of the other panels, over the range $8\leq m \leq 15$. Both the full frequency resolution (black) and the smoothed (red line) spectrum are shown. The blue vertical lines show the central frequency (solid line) and the leakage locations (dashed). The ``overtone" appears as a peak in the averaged spectrum with the leakage on both sides appearing weaker or absent. This suggests that this feature is retrograde and not a component of the prograde magnetic activity.}
    \label{fig.overtones}
\end{figure*}

Unlike the Rossby or \HFR modes, the presence of latitudinal overtones in the observed spectra is not clearly discernible above the noise. However, in the context of the results of \citet{blume_etal_2024}, {\cite{jain_etal_2024}} and \citet{bhattacharya_hanasoge_2023}, an examination of the $\ell=m+2$ channel, which is symmetric in latitude, suggests the presence of enhanced power at frequencies approximately 60\% greater than the \HFR modes. Panels \textbf{A} and \textbf{B} of Fig.~\ref{fig.overtones} compare the antisymmetric $\ell=m+1$ spectra to the symmetric $\ell=m+2$ spectra. We highlight the region of interest with the white dashed line, as well as the Rossby (red dashed) and \HFR (cyan dashed) dispersions for context. Due to differential rotation, the latitudinal eigenfunctions of the Rossby modes are not accurately captured by singular spherical harmonics, implying the presence of a strong signal in the $\ell=m+2$ channel along the Rossby dispersion. The \HFR ``overtone" feature appears as a marginally enhanced power above the noise in the $\ell=m+2$ channel. This feature is very weak, appearing only when smoothing and thus determining an appropriate dispersion is difficult. Examination of the entire spectra makes it difficult to assert if this signal is prograde or retrograde, much like the divergence signal analyzed already. Following the same $m$ averaging used in the previous section, we find a weak but significant ridge of power (panel \textbf{C}) that appears to be retrograde. The signal-to-noise ratio of the mode-averaged spectra is very poor ($\sim0.3$), though given its proximity to the frequencies of the overtones in the models, it is tempting to associate this tenuous detection with the modes predicted by models.

\section{Numerical Modeling and Analysis}\label{sec:numerical}
Thus far, we have confirmed from the observations that the \HFR modes possess a poloidal-flow component in the near-surface layers of the Sun. This characteristic has been suggested by all numerical \HFR studies \citep{triana_etal_2023,bhattacharya_hanasoge_2023,bekki_2024,blume_etal_2024}, although deeper in the interior of the convection zone. It may be tempting to directly compare the observations with the models in order to make further inferences about the solar interior. However, there are still a number of caveats with the modelling that ought to be considered. Here, we utilize the linear spectral solver of \citep{bhattacharya_hanasoge_2023} to explore the closeness of the model \HFR modes to observations, and areas that still need improvement.

\begin{figure}
    \centering
    \includegraphics[width=\linewidth]{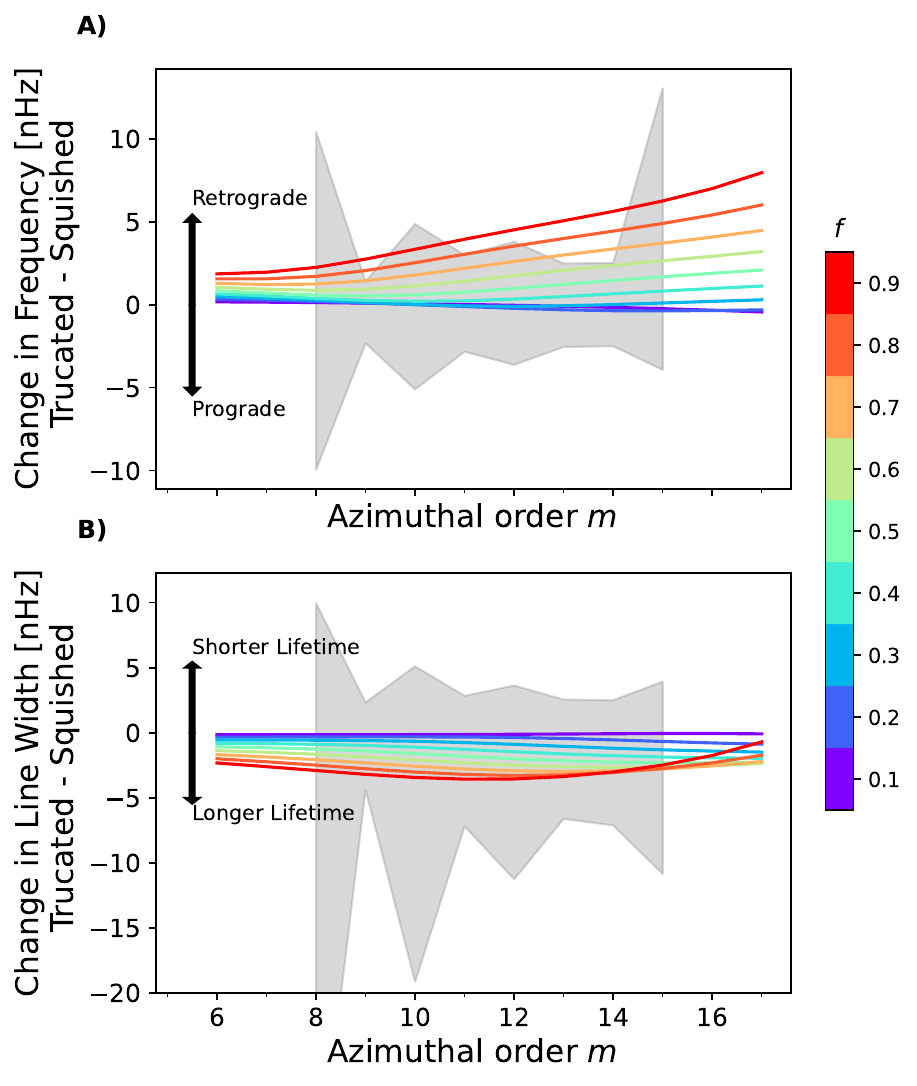}
    \caption{The difference in the \HFR mode frequency (panel \textbf{A}) and line width (panel \textbf{B}) between the commonly used truncated rotation profile and a model where the entire rotation profile is squished into the model domain. The different line colors signify the fractional coefficient $f$ applied to the differential rotation (in the co-rotating frame) to illustrate the gradual change in the model parameters as they shift from uniform rotation ($f=0$) to the solar differential rotation ($f=1$). The shaded regions signify the error estimates on the mode parameters derived from the observations used in this study.}
    \label{fig.paramsRot}
\end{figure}

\begin{figure}
    \centering
    \includegraphics[width=\linewidth]{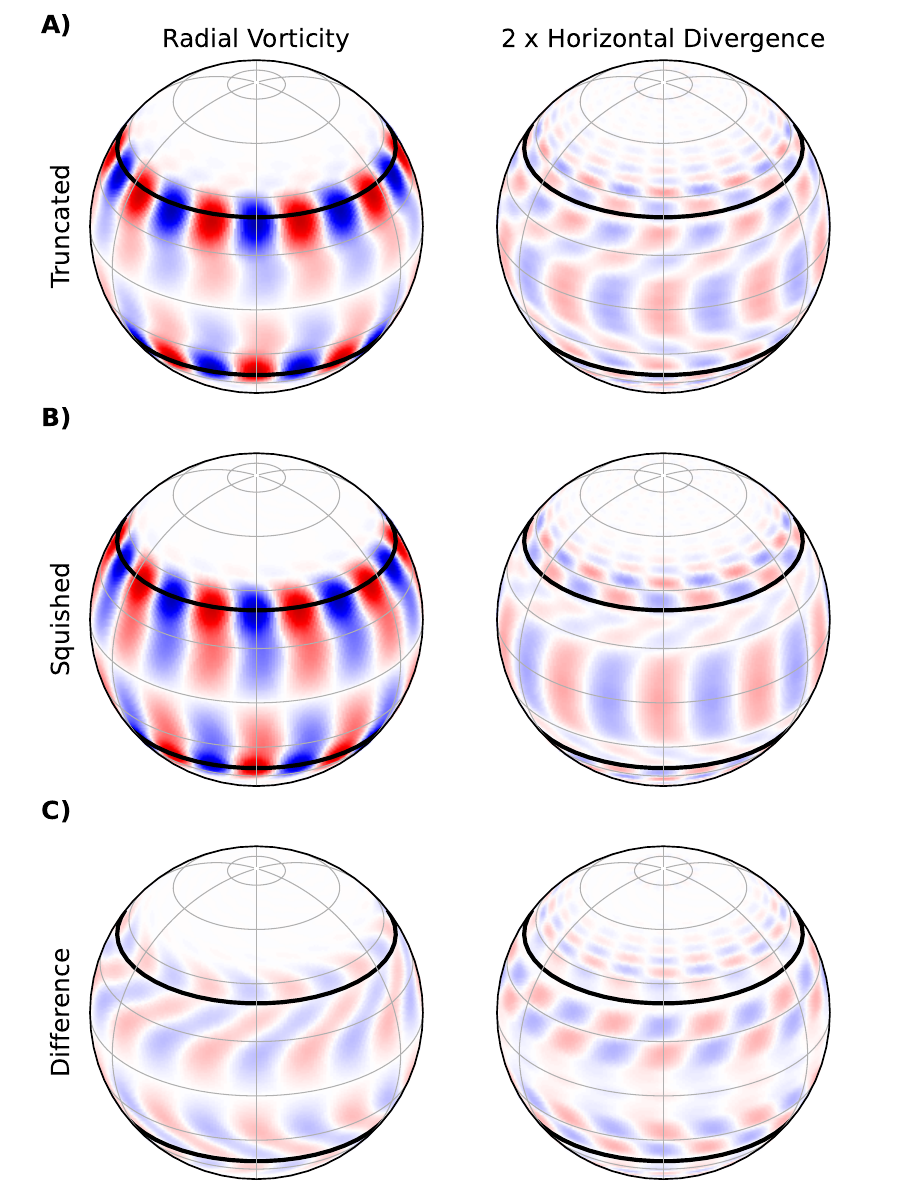}
    \caption{Surface eigenfunctions of the $m=10$ \HFR mode's radial vorticity (left column) and the horizontal divergence (right column). Row \textbf{A} shows the two eigenfunctions in the case of the truncated model that ignores the near-surface rotation and stratification. Row \textbf{B} shows the eigenfunctions in the case where the near-surface rotation is squished into the model domain. Row \textbf{C} shows the difference (truncated - squished) between the eigenfunctions. The right hand column has been multiplied by 2 for visualization on the same color scale.}
    \label{fig.horeig}
\end{figure}

The outer boundary of linear stratified models is chosen to be located just below the solar near-surface shear layer  \citep[0.985~R$_\odot$,][]{bekki_etal_2022b,bhattacharya_hanasoge_2023}. The near surface sees a rapid change in stratification (six orders of magnitude in density), making this region very challenging to resolve numerically. It is thus a common practice to excise this region from the model domain, which we refer to here as the {\it truncated} model. However, most seismic modes used in local helioseismic observations, especially here, propagate primarily within this region. 
\citet{bhattacharya_etal_2024} attempted to account for at least the rotational component of the near-surface shear layer (ignoring the steep density stratification) by squishing the observed rotational profile into the computational domain, which we refer to as the {\it squished} model. Figure~\ref{fig.paramsRot} shows how the \HFR mode spectrum differs between the {\it truncated} and {\it squished} models. The presence of the near-surface rotational shear results in more retrograde oscillation frequencies (greater negative values), with longer lifetimes (smaller line widths). Comparison with observational errors shows that these changes are within or comparable to the error estimates. In Fig.~\ref{fig.horeig} we compare the surface horizontal eigenfunctions for the \HFR mode at $m=10$ for these two cases. We find that the surface radial vorticity is qualitatively similar between the {\it truncated} and {\it squished} models, the difference being approximately a fifth of the amplitude of the eigenfunctions. However, for the horizontal divergence, there is a more significant change in the eigenfunction due to the inclusion of the near-surface shear with the difference between the cases having an amplitude comparable to the eigenfunctions.
The ability of the models to reasonably capture most of the \HFR spectrum (frequency and life times) suggests that the inclusion of this region may not drastically alter current inferences that rely only on these parameters \citep[e.g.][]{bekki_2024}. However, future efforts that rely on the eigenfunctions will need to consider the role of the near-surface rotation. The consequence on the \HFR spectrum or eigenfunctions due to near-surface density and sound speed stratification remains an open question.

\begin{figure}
    \centering
    \includegraphics[width=0.9\linewidth]{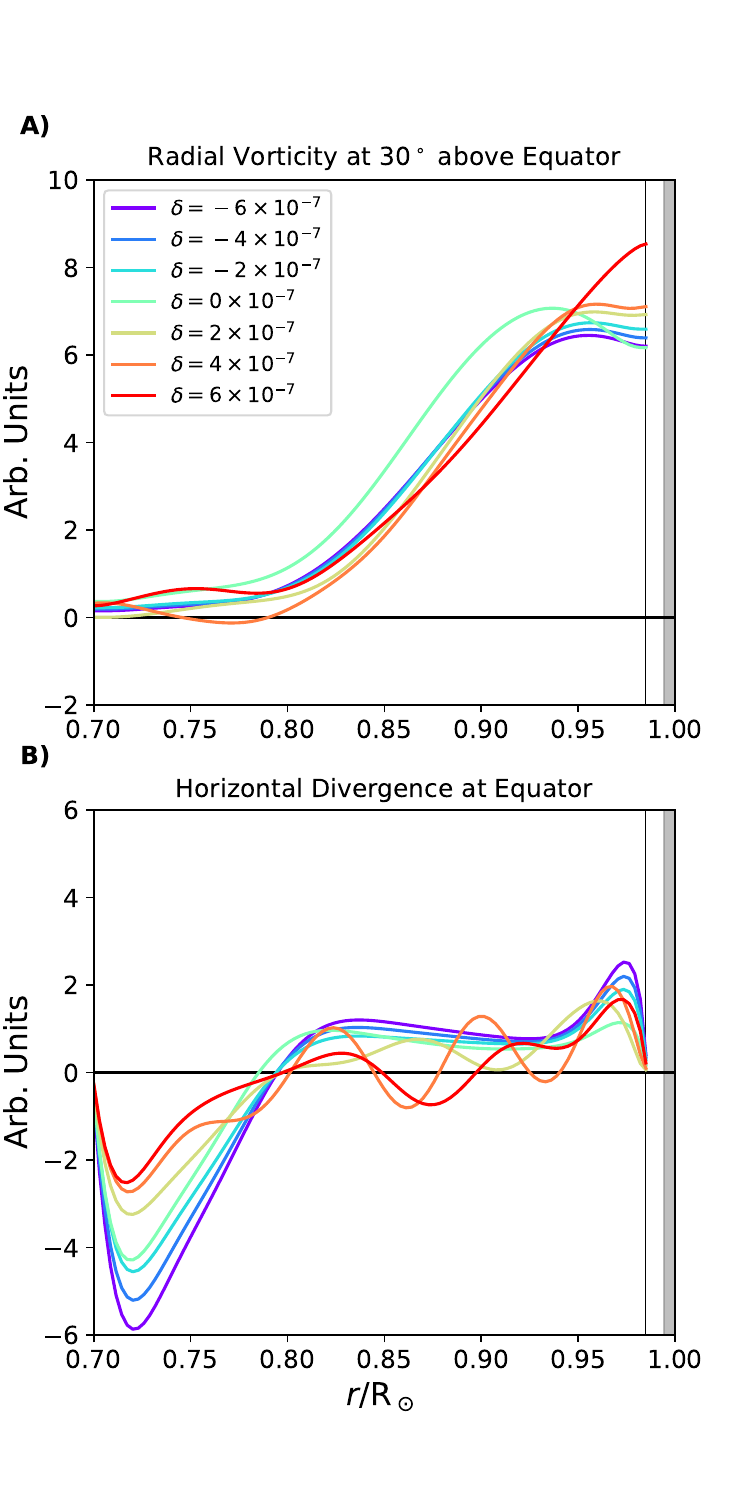}
    \caption{ Slices of the radial vorticity (\textbf{A}) and horizontal divergence (\textbf{B}) for the $m=10$ \HFR mode. The slices are taken at the latitude of the maximal power, which is $30^\circ$ and $0^\circ$ above the equator for the vorticity and divergence, respectively. The black vertical line signifies the boundary of the model domain. The shaded grey region on the right is an estimate of the mode cavity of the f and p modes used in the observational analysis. Different line colors signify different superadiabatic temperature gradients $\delta$, a free and uniform parameter of the model, which has been shown to greatly affect the \HFR modes properties \citep{bekki_2024}.}
    \label{fig.slices}
\end{figure}

Current linear models also apply an impenetrable boundary condition at their surface (0.985R$_\odot$), such that there is no radial flow $u_r$ across the boundary. The equation for mass conservation connects the horizontal divergence of the inertial-mode motions to the radial derivative of the radial velocity. Setting $u_r$ to zero at the boundary influences the radial derivatives and so the horizontal divergence is affected; these effects are transmitted through the bulk of the simulation domain. In Fig.~\ref{fig.slices}, we plot a slice through the radial-vorticity and horizontal-divergence eigenfunctions (at the latitudes of their respective maxima). In general, the horizontal-divergence eigenfunctions undergo rapid decay in their profiles near the top and bottom boundaries, while the vorticity appears comparatively unaffected by the presence of the boundaries. It is important to determine how shifting the outer boundary to R$_\odot$ would influence the mode eigenfunctions. Additionally, the effect of other choices for surface boundary conditions should also be investigated, since there is no reason that inertial modes should be limited to the solar interior.

Despite our cautious tone, numerical models have done remarkably well in capturing solar inertial mode physics. While the horizontal divergence eigenfunction is sensitive to the boundary condition and rotation, the radial vorticity eigenfunction and mode parameters appear more robust. The relative amplitude of the horizontal divergence to the radial vorticity from our model is $\sim 0.15$ which is approximately {3} times smaller than the observations where the ratio is is $\sim 0.5$ . While tuning the model parameters and boundary conditions may bring about a closer agreement, this is already to be regarded as evidence that the models are in good stead for future inferences.

\section{Conclusions}\label{sec:conclusion}

The discovery of inertial modes raises the possibility of `inertial mode seismology': utilizing inertial modes to infer the internal properties of the Sun. In order to fully realize the promise of this technique, we must first accurately characterize the observed inertial modes. Numerical studies have not only reported on the presence of \HFR modes in simulations of stratified rotating spheres, but also suggested that \HFR waves possess a significant non-toroidal (radial) flow component deep within the solar convection zone. Using 13 years of HMI $5^\circ$ ring-diagram fit parameters, we have constructed very-shallow depth-averaged ({$\sim0.6\%$ in solar radius}) flow maps. We report the presence of a signal in the power spectrum of sectoral horizontal divergence that coincides with previously reported frequencies. Furthermore, we also report the tenuous detection of an overtone of the \HFR modes, as predicted by \citet{bhattacharya_hanasoge_2023} and \citet{blume_etal_2024}. This feature is retrograde and appears in the equatorially symmetric $\ell=m+2$ channels, suggesting it has two nodes in the radial vorticity eigenfunction.

With equatorially anti-symmetric radial vorticity and symmetric horizontal divergence, the \HFR modes are reminiscent of convective thermal Rossby modes \citep{busse_1970,gilman_1975}, also referred to as banana cells or Busse columns in the literature. Thermal Rossby are purely prograde according to all studies, which discounts the possibility that the \HFR modes are the same as these important but observationally elusive modes. \citet{blume_etal_2024} and {\citet{jain_etal_2024} have both} suggested that the \HFR modes, and their latitudinal overtones, may be the retrograde branch of mixed modes with prograde thermal Rossby modes. We explored the spectra for any possible cross over of the \HFR modes at $m=0$, which is indicative of mixed modes, but we found no prograde branches. However, given the high retrograde frequency of the \HFR modes, it is likely that the crossover would occur near or above the rotation rate $-\Omega_{\textrm{ref}}$, which results in complications with leakage, making this a non-trivial exercise. Given that the models have been accurate so far in both the \HFR divergence and latitudinal overtones, it is possible that the reason for not observing a prograde branch of the mixed modes may be related to the same reason that pure thermal Rossby modes have eluded observers for decades.

This work represents progress derived from a dialogue between numerical modelling and observations of solar inertial modes. Specifically, models have been ahead of the present observations, predicting the existence of the radial flow component of \HFR modes as well as possible latitudinal overtones. Further effort is still required to build trust in the numerically derived eigenfunctions; however, the current agreement between observations and modelling suggests that we are one step closer to fully utilizing inertial modes to explore the deep interior of the Sun.

\section*{Acknowledgments}
This research is based on work supported by Tamkeen under the NYU Abu Dhabi Research Institute grant CASS. SH acknowledges funding from the Department of Atomic Energy, India. SH and KRS also acknowledge support from the King Abdullah University of Science and Technology (KAUST) Office of Sponsored Research (OSR) under award OSR-CRG2020-4342.
The data used in this analysis is courtesy of NASA/SDO and the HMI science teams. This research was carried out on the High Performance Computing resources at New York University Abu Dhabi.

\software{The observational codes\footnote{\url{https://github.com/cshanson/Solar-Inertial-Modes}} and numerical codes\footnote{\url{https://github.com/jishnub/RossbyWaveSpectrum.jl}} used here have been made freely available on Github, under the MIT license. The observational codes are in the Python language and the numerical codes are in Julia language.}

\newpage
\appendix
\renewcommand\thefigure{\thesection.\arabic{figure}}
\setcounter{figure}{0}    

\section{Analysis of sectoral Rossby Waves}

\begin{figure}[!h]
    \centering
    \includegraphics[width=\linewidth]{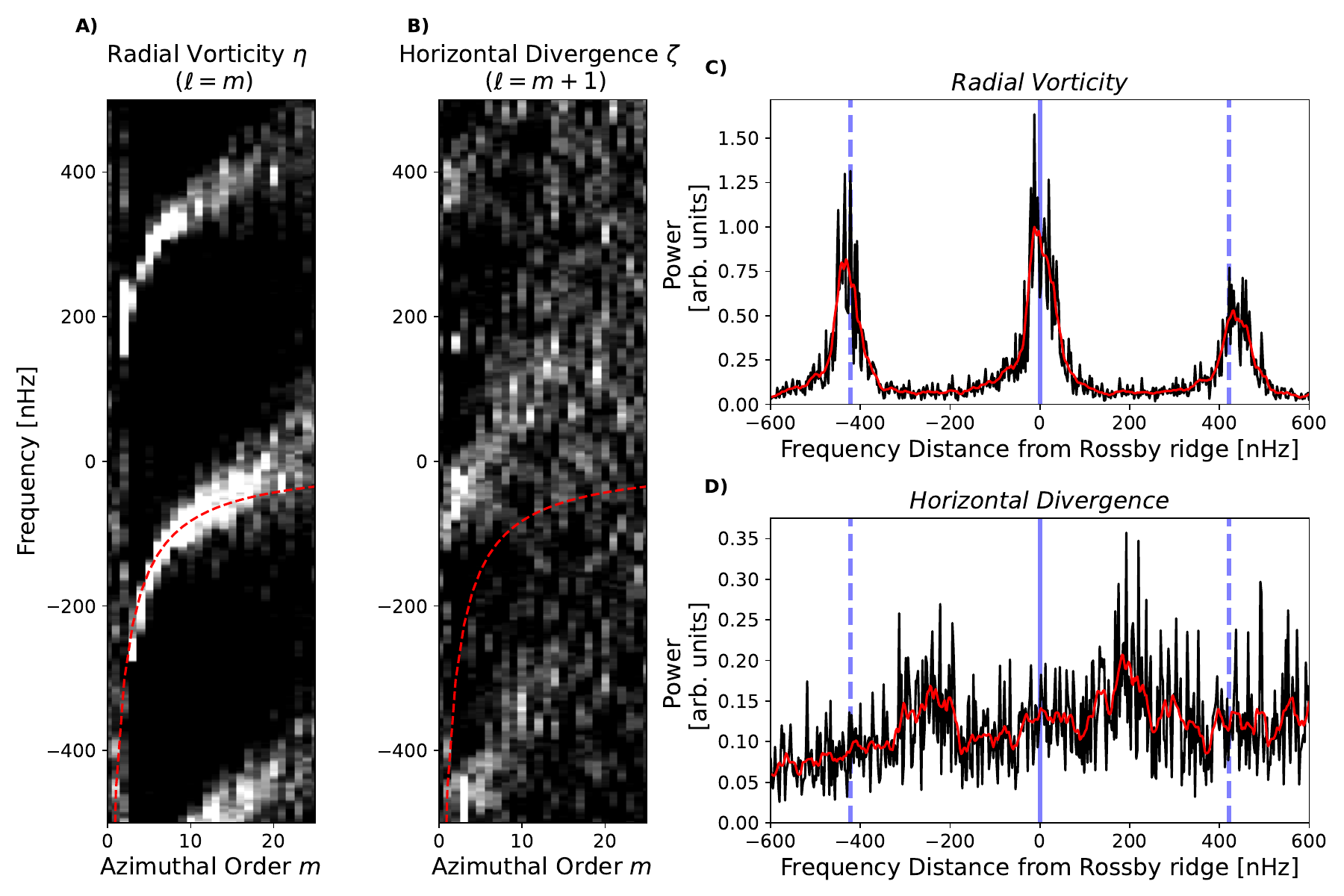}
    \caption{Radial vorticity ($\ell=m$) and horizontal divergence ($\ell=m+1$) of the solar Rossby waves. Same figure notation as Fig.~\ref{fig.spectra}. In panels \textbf{C} and \textbf{D} the centered frequencies are based on the measured dispersion of \citet{loeptien_etal_2018}. Unlike the \HFR modes the Rossby waves have no horizontal divergence signal, above the noise, at the surface.}
    \label{fig.spectra_RH}
\end{figure}



\bibliography{References_DONTDEL}{}
\bibliographystyle{aasjournal}

\end{document}